\documentstyle[12pt,epsf,fleqn]{modif}

\def \be {\begin{equation}}
\def \ee {\end{equation}}

\begin{document}

\begin{frontmatter}
\title{Background and Technical Studies for GENIUS as a Dark Matter Experiment}
\noindent
\author{L. Baudis, G. Heusser, B. Majorovits, Y. Ramachers, }
\author{H. Strecker and H. V. Klapdor--Kleingrothaus}
\address{Max--Planck--Institut f\"ur Kernphysik, Heidelberg, Germany}

\begin{abstract}

The GENIUS project is a proposal for a new dark matter detector, 
with an increased sensitivity of three orders of magnitude relative
to existing direct dark matter detection experiments.
We performed a technical study and calculated the main background
sources for the relevant energy region in a detailed detector geometry. 
The achieved overall background level and detector performance confirm 
the outstanding potential of GENIUS as a powerful tool for the direct search of
WIMPs in our Galaxy.
\end{abstract}
\end{frontmatter}

\section{Introduction}

GENIUS (GErmanium in liquid NItrogen Underground Setup) is a proposal
for operating a large amount of 'naked' Ge detectors in liquid nitrogen 
for dark matter and alternatively $\beta\beta$--decay researches 
\cite{ringb,hirsch,hellmig97,klap}, with an improved sensitivity 
of three orders of magnitude relative to present experiments. 
The idea to operate ionization HPGe detectors 
directly in liquid nitrogen has been already discussed in \cite{heusser}. 
In \cite{hellklap} it has been shown,
that the detectors work reliably under such conditions. 
If the detector performance in liquid nitrogen is unaltered or even better
than for usual Ge detectors, the liquid can act both as a cooling
medium and as an effective shielding
against external radioactivity, since liquid nitrogen can be processed
to a very high purity. The proposed scale of the experiment 
would be a nitrogen tank of about 12 m diameter and 12 m height 
which contains in its dark matter version 100 kg (40 detectors) of natural
germanium suspended in its center. 
The optimal location would be the Gran Sasso Underground Laboratory.  

To cover large parts of the MSSM parameter space, relevant for the
detection of neutralinos as the dark matter candidate \cite{jkg96,bedny1}, 
a maximum background
level of 10$^{-2}$ counts/(kg y keV) in the energy region between
0--100 keV has  to be achieved. 
This means a very large further background reduction
in comparison to our recent best result (20.82 counts/(kg y keV))
\cite{prl} with an enriched detector of the Heidelberg--Moscow
experiment \cite{hdmo} and to all other running dark matter
experiments (for a review see \cite{klap}).  

\section{Technical study}

To demonstrate the feasibility of operating Ge detectors in liquid nitrogen,
instead of in a vacuum--tight cryostat system \cite{kno89}, 
a first experiment has been successfully performed in the low level
laboratory in Heidelberg with one naked p--type Ge crystal immersed in a 50
l dewar \cite{hellmig97}. 
Already in this attempt we could not see any deterioration in the
detector performance relative to our conventionally operated detectors.

In a second phase the goal was to look for possible interferences between 
two or more naked Ge crystals, to test different cable lengths between
FETs and crystals and to design and test a preliminary holder system
of high molecular polyethylene.
We started a technical study operating three germanium detectors on a
common plastic holder system inside liquid nitrogen. 
All crystals were of p--type and weighted about 300 g each.

\begin{figure}[t]
\vspace*{12cm}
\epsfxsize12cm
\centerline{\epsffile{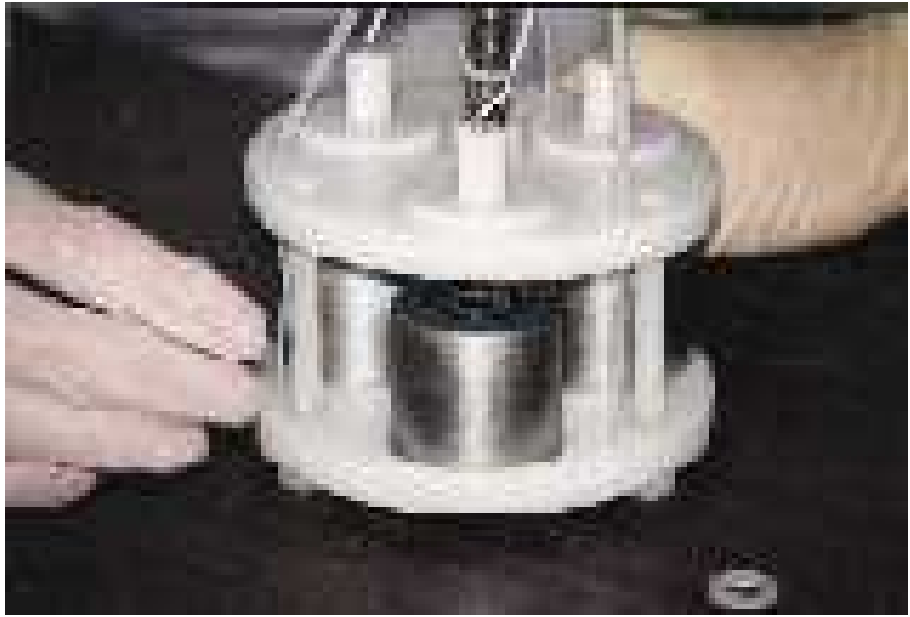}}
\caption{The three--crystal holder-system with germanium crystals mounted 
shortly before cooling. Some crystal-to-FET cables can be seen.}
\label{canberra2}
\end{figure}

A picture of the three--crystal holder--system can be seen in 
figure ~\ref{canberra2}. 
Two thin polyethylene plates (1 cm thick) are used to fix the contacts 
to the crystals. The FETs are placed close to the liquid nitrogen 
surface but kept inside. Cables having three different lengths (2, 4 and 
6 m) connect the three crystals to their FETs. 

The main purpose of the experiment was to test the behaviour of the
crystals in the low energy region: energy resolution, energy
threshold and possible signs of microphonic events caused by nitrogen boiling. 
The general performance 
of the crystals is as stable as already seen with a single detector inside 
liquid nitrogen. We couldn't observe any cross talk using only p--type
detectors (same polarity for the HV-bias), since cross talk signals have the wrong 
polarity and are filtered by the amplifier.

\begin{figure}[h]
\epsfxsize12cm
\centerline{\epsffile{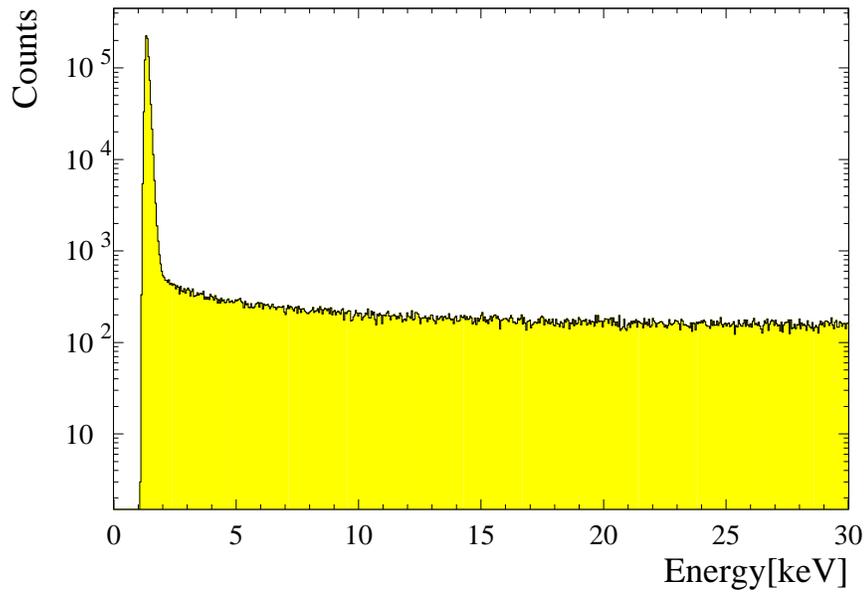}}
\caption{Background spectrum of a naked, unshielded Ge crystal in liquid nitrogen. 
Note the low energy threshold of 2 keV of the detector.}
\label{topfgs_back}
\end{figure}

\begin{figure}[t]
\epsfxsize12cm
\centerline{\epsffile{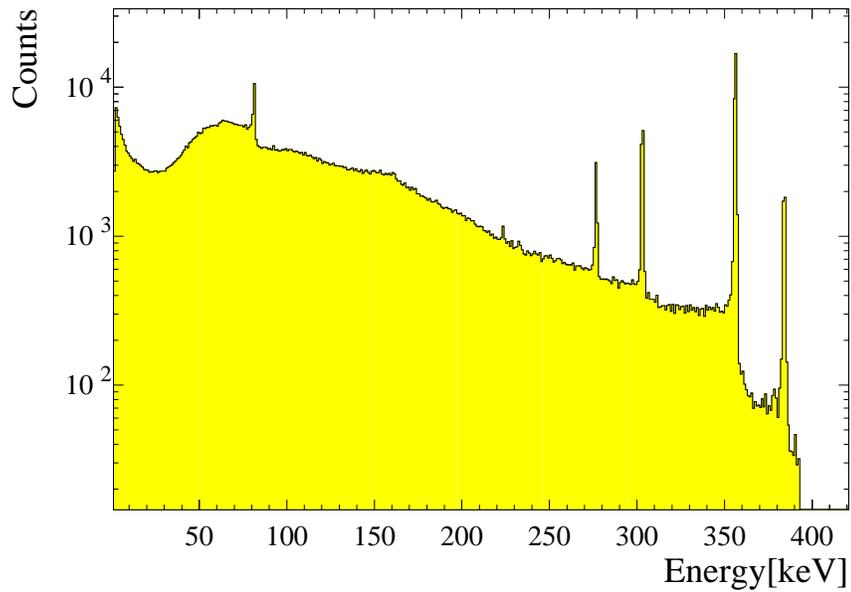}}
\caption{Calibration $^{133}$Ba spectrum of a naked Ge crystal in liquid nitrogen. 
The energy resolution is 1 keV at 300 keV.}
\label{baspec}
\end{figure}

Figure \ref{topfgs_back} shows a background spectrum and
figure \ref{baspec} a $^{133}$Ba calibration spectrum of one of the naked Ge
detectors in liquid nitrogen. The
cable length between detector and FET was 6 m (winded up in loops). 
We achieved ab energy energy resolution of 1.0 keV at 300 keV and a threshold of 2 keV. 
No microphonic events due to nitrogen boiling beyond 2 keV could be detected. 
We conclude that the performance of the Ge detectors is as good (or even
better) as for conventionally operated crystals, even with 6 m cable
lengths between crystal and FET. 

\section{Background considerations}

The aim of the GENIUS experiment is to reach the extremely low background
level of 0.01 events/(kg y keV) in the energy region below 100 keV.
To show that this is indeed achievable, we performed detailed Monte Carlo
simulations and calculations of all the relevant background sources.
The sources of background can be divided into external and internal ones.
External background is generated by events originating from outside
the liquid shielding, such as photons and neutrons from the Gran Sasso
rock, muon interactions and muon induced activities.
Internal background arises from residual impurities in the liquid
nitrogen, in the steel vessel, in the crystal holder system, in the Ge crystals
themselves and from activation of both liquid nitrogen and Ge crystals
at the Earths surface.

For the simulation of muon showers, the external photon flux
and the radioactive decay chains we used the GEANT3.21 package 
\cite{geant} extended for nuclear decays \cite{mueller}.
This version had already successfully been tested in establishing a
quantitative background model for the Heidelberg--Moscow experiment
\cite{hdmo}. 

In a first crude estimation the influence of the components expected
to be dominant were studied \cite{hellmig97,hellklap}. 
In the simulation the setup consisted of a tank, 9 m
in height and 9 m in diameter, with 288 naked enriched germanium
crystals of 3.6 kg each, positioned in its center (this simulation has 
been performed in view of GENIUS as a neutrinoless
double beta decay experiment with 1 ton of enriched $^{76}$Ge). 
From the background sources mentioned above, only the activities in
the liquid nitrogen, steel vessel and muon showers were simulated.

For a more detailed study of the sensitivity of GENIUS in a first experimental phase,
further simulations with a new, more accurate geometry have been
carried out \cite{bela98}.
The vessel is surrounded by a 2 m thick polyethylene-foam isolation, which 
is held by two 2 mm thick steel layers 
(constructional data from Messer--Griesheim).
The nitrogen shielding is given by a cylindrical geometry of 12 m
diameter and 12 m height with the crystals in its center.
The simulated setup consisted of 5 natural Ge detectors 
integrated into a holder system of high molecular polyethylene. 
In comparison to the first simulation, the active mass reduces to
approximately 14 kg and the anticoincidence of the detectors \cite{hellmig97}
will obviously be less effective.

\subsection{External background}

\subsubsection{Photon and neutron flux from the surroundings}

We simulated the influence of the photon flux with energies between
0 -- 3 MeV measured in hall C of the Gran Sasso laboratory
\cite{arpesella92}. This measurement is in good agreement with photon flux
calculations by the Borexino Collaboration \cite{borexprop}.  
The main contributions are given in table \ref{photonGS}.
With a diameter of 12 m for the liquid shielding, we obtain a count
rate of 4$\times$10$^{-3}$ counts/(kg y keV) in the energy region 0 -- 100 keV. 

The influence of the measured neutron flux in the Gran Sasso
laboratory \cite{arp} was estimated.  
The 2 m polyethylene foam isolation ($\rho$ = 0.03 g cm$^{-3}$) reduces
the neutron flux for energies below 1 keV by more than 5 orders of
magnitude. Only about 3\% of neutrons with energies between 1 keV
and 2.5 MeV will pass the polyethylene isolation, whereas for energies
between 2.5 and 15 MeV the overall flux is reduced by about 40\%.
The neutron flux reaching the tank can be reduced by another
two orders of magnitude by doping the polyethylene foam isolation with
about 1.4 t of boron.  
The flux of the $^7$Li deexcitation gamma rays from
the reaction  n + $^{10}$B$\rightarrow \alpha $ + $^7$Li*, with an 
energy of 0.48 MeV,  would be too low
to reach the inner part of the liquid shielding.
After the first meter of liquid nitrogen the total flux is reduced
by another 4--5 orders of magnitude, therefore we simulated the neutron capture
reactions randomly distributed in the first meter of the nitrogen shielding.   

With the assumption that all neutrons reaching the nitrogen are thermalized and  
captured by the reactions $^{14}$N(n,p)$^{14}$C$^{*}$ and 
$^{14}$N(n,$\gamma$)$^{15}$N$^{*}$, 
a total of 4.4$\times$10$^{7}$ 
neutron capture reactions per year have to be taken into account. 
The relevant contribution to the background comes from the
deexcitation of the $^{14}$C$^{*}$ and $^{15}$N$^{*}$ nuclei.
The contribution of the $\beta$--decay of $^{14}$C nuclei in the liquid nitrogen 
is negligible, since only low energy electrons (E$_{\beta max}$ = 156
keV) are emitted and the decay probability is very low due to the long 
half life (10$^{-4}$ per year).

Using the assumptions from above, 
the mean count rate in the low--energy region due to neutron capture
reactions would be about 4$\times10^{-4}$ counts/(kg y keV). 

\subsubsection{Activities induced by muons}

The muon flux in the Gran Sasso laboratory was measured to be
$\phi$$_{\mu}$=2.3$\times$10$^{-4}$ m$^{-2}$s$^{-1}$ with a mean
energy of $\bar{E}_{\mu}$=200 GeV \cite{arpesella92}.

We simulated the effect of muon induced showers in the liquid
nitrogen. With the aid of a muon veto in form of scintillators or gas counters
on top of the tank, the total induced background can be drastically reduced.
Here we assumed a veto efficiency of 96\% as measured in a more
shallow laboratory \cite{heusser91}. The count rate reached in the low--energy
region is about 2$\times$10$^{-3}$ counts/(kg y keV).
It can be further improved using the anticoincidence power of the 40
Ge detectors of the future dark matter setup among each other.

Besides muon showers, we have to consider muon induced nuclear 
disintegration and interactions due to secondary neutrons generated in the above
reactions. 

\subsubsection*{Neutrons generated by cosmic muons}

The muon induced production of neutrons can be approximated by
A$_{n} \sim$ 3.2$\times$10$^{-4}$ (g$^{-1}$ cm$^{2}$), due to the
$<E>^{0.75}$ dependence of the number of neutrons on the mean muon 
energy \cite{myonneutrons}. 
With the geometry of the tank h = 12 m, r = 6 m, the density of
nitrogen $\rho$ = 0.808 g/cm$^{3}$ and the cited flux, a mean
production rate of $\phi _{n\mu}$ = 2.5$\times$10$^{5}$ neutrons/year in
the whole vessel is obtained.
Table \ref{neutrons} gives the neutron induced reactions
in the liquid nitrogen for neutron energies $<$ 20 MeV (based on
all reactions found in \cite{McLane88}).

All of the produced nuclides are stable or short lived with the exception
of $^{14}$C and $^{13}$N. The contribution of gammas from the excited
$^{14}$C$^*$ nucleus corresponds to 10$^{-3}$ counts/(kg y keV) between
0 -- 100 keV. The contribution from the $\beta^{-}$-- particles
with E$_{max}$ = 0.16 MeV is negligible due to the low decay probability
of  $^{14}$C. The production rate of $^{13}$N is 
1$\times$10$^6$ atoms per year in the whole tank. 
From 10$^6$ simulated positrons with E$_{max}$ = 1.2 MeV ($\beta
^+$-decay), corresponding to an 
exposure of about 1 year, only one event could be observed in the detectors. 
Therefore, the contribution of $^{13}$N to the background will be negligible.

\subsubsection*{Negative muon capture}

A negative muon stopped in the liquid shielding can be captured by a 
nitrogen nucleus, leading to one of the reactions that are listed in 
table ~\ref{myonspall}.
Estimations of the number of stopping muons in the nitrogen tank
\cite{bergamasco82,gaisser,lohman} lead to 86 stopped muons per day.
The rates of decaying and captured muons are shown in table
\ref{mcapture}.
 
The derived production rates \cite{charalambus} for the various isotopes
are listed in table \ref{myonspall}.
Only the isotopes 
$^{14}$C, $^{10}$Be, $^{11}$C and $^{10}$C
can not be discriminated by 
muon anticoincidence, since their individual lifetime is too long.
$^{14}$C and $^{10}$Be will not be seen in our detector due to
their very low decay probabilities (10$^{-4}$ and 10$^{-10}$ per year) 
and low production rates.
The contribution of $^{10}$C and $^{11}$C,
with a production rate of 117 atoms/year and 292 atoms/year,
respectively, in the whole nitrogen tank, will be negligible.
The gamma rays from the excited nuclei produced in the reactions can
be discriminated by the anticoincidence shielding.

\subsubsection*{Inelastic muon scattering}

Another way of producing radioactive isotopes in the liquid
shielding are electromagnetic nuclear reactions of muons through
inelastic scattering off nitrogen nuclei: $\mu$ + N $\rightarrow$ $\mu'
$ + X$^*$. The only resulting isotopes with half lifes $>$ 1s
are $^{14}$N($\gamma$,n)$^{13}$N, with T$_{1/2}$=9.96 m and 
$^{14}$N($\gamma$,tn)$^{10}$C, with T$_{1/2}$=19.3 s. 
The production rate per day for one isotope can be written as
\cite{OConnell88}:
R(d$^{-1}$)=
6$\times$10$^{-2}\phi_{\mu}$(d$^{-1}$m$^{-2}$)N$_{T}$(kt)$\sigma_{\mu}$($\mu$b)/A,
where $\phi _{\mu}$ is the flux of muons on the detector, N$_{T}$ is the
number of target nuclei, $\sigma_{\mu}$ the reaction cross section and 
A the atomic weight of the target nucleus. For our detector this
yields R(y$^{-1}$)= 45$\times$$\sigma_{\mu}$($\mu$b).
With typical reaction cross sections of a few hundred $\mu$b \cite{OConnell88,Napoli73},
we obtain a production rate of (5--10)$\times$10$^3$ atoms per year.
A simulation of an activation time of ten years for both isotopes
yields negligible count rates in comparison to contributions
from other background components.

\subsection{Internal background}

\subsubsection{Intrinsic impurities in the nitrogen shielding, Ge crystals,
  holder system and steel vessel}

The assumed impurity levels for the liquid nitrogen 
are listed in table ~\ref{forderung}.
For the $^{238}$U and $^{232}$Th decay chains they have already been
reached by the Borexino experiment \cite{borexino} for their liquid scintillator. 
Due to the very high cleaning efficiency of fractional distillation, 
it is conservative to assume
that these requirements will also be fulfilled for liquid nitrogen.
The $^{238}$U and $^{232}$Th decay chains were simulated under the assumption that 
the chains are in secular equilibrium. 
The count rate due to $^{238}$U, $^{232}$Th and $^{40}$K
contaminations of the liquid nitrogen is about 3.1$\times 10^{-3}$ in the
energy region below 100 keV. 

Measurements indicate that the $^{222}$Rn contamination of freshly
produced liquid nitrogen is in the range of  10 mBq/m$^{3}$.
After about a month it is reduced to 100 $\mu$Bq/m$^{3}$ (T$_{1/2}$ =
3.8 days). 
Such a level could be maintained if the evaporated nitrogen is always
replaced by Rn--pure nitrogen, previously stored in the underground. 
Surface emanations are reduced to a negligible level for cooled
surfaces in direct contact with the liquid nitrogen.
Non cooled surfaces have to be highly contamination free from $^{222}$Rn.

The mean count rate
from the contamination of $\;^{222}$Rn in the interesting
region below 100 keV is  10$^{-3}$ counts/(kg y keV) assuming an
activity of 100 $\mu$Bq/m$^{3}$ in the liquid nitrogen. 

For the intrinsic impurity concentration in Ge crystals we can
give only upper limits from measurements with the detectors of the
Heidelberg--Moscow experiment. 
We see a clear $\alpha$--peak in two of the enriched detectors at 5.305
MeV, and an indication for the same peak in two other detectors. It
originates from the decay of $^{210}$Po (which decays with 99\%
through an $\alpha$--decay to $^{206}$Pb) and is a sign for a $^{210}$Pb
contamination of the detectors. However, it is very unlikely that the
contamination is located inside the Ge-crystals, most probably it is located on
the crystals surface at the inner contact.    

Using three Ge detectors, we
derive an upper limit at 90\% CL (after 19 kg y counting statistics)
of 1.8$\times$10$^{-15}$g/g for $^{238}$U and 5.7$\times$10$^{-15}$g/g 
for $^{232}$Th. 
Assuming these impurity concentrations, our
simulations yield a count rate of about 10$^{-2}$ counts/(kg y keV)
for both $^{238}$U and $^{232}$Th decay chains.
It is however secure to assume that these upper
limits are very conservative and that the true contamination
of HPGe is much lower. Special attention will have to be paid in order
to avoid surface contaminations of the crystals.     


An important open factor in the background spectrum 
is the effect of the holder-system. 
For the simulation we assumed the possibility to obtain a polyethylene
with an impurity concentration of 10$^{-13}$g/g for the U/Th decay
chains (this is a factor of 100 higher than the values reached for the
organic liquid-scintillator by the Borexino collaboration \cite{borexino}). 
Encouraging are the results already achieved by the SNO experiment
\cite{sno}, which developed an acrylic with current limits on 
$^{232}$Th and $^{238}$U contamination of 10$^{-12}$g/g.
Since it is not yet sure that such a low contamination level will be 
reached for polyethylene, we are currently testing also other materials.

Assuming the above impurity level with the simulated geometry a count
rate of $\sim$ 
8$\times$10$^{-4}$ counts/(kg y keV) in  the energy region below 100
keV from this component is reached. This result will be further improved by
development of a new holder design using a minimized amount of material. 

For the steel vessel an impurity concentration of 1$\times$10$^{-8}$
g/g for U/Th was assumed (Borexino measured 5$\times$10$^{-9}$ g/g
\cite{borexprop}). The contribution in the energy region 0 -- 100 keV is
1.5 $\times$ 10$^{-5}$ counts/(kg y keV). Assuming an equal
contamination for the polyethylene foam isolation as for steel, 
the contribution of both to the background is negligible. 

\subsubsection{Cosmic activation of the germanium crystals}

We have estimated the cosmogenic production rates of radioisotopes 
in the germanium crystals with the $\Sigma$ programme \cite{JensB}.
The programme was developed to calculate cosmogenic activations of
natural germanium, enriched germanium and copper. 
It was demonstrated that it can  reproduce the measured cosmogenic
activity in the Heidelberg--Moscow experiment \cite{hdmo} 
within about a factor of two \cite{BerndM}.  

Assuming a production and transportation time of 10 days at sea level
for the natural Ge detectors, and a deactivation time of three years,
we obtain the radioisotope concentrations listed in table \ref{ge_cosmo}.
All other produced radionuclides have much smaller activities due to their shorter
half lifes. 

\begin{figure}[h]
\epsfysize=10cm
\centerline{\epsffile{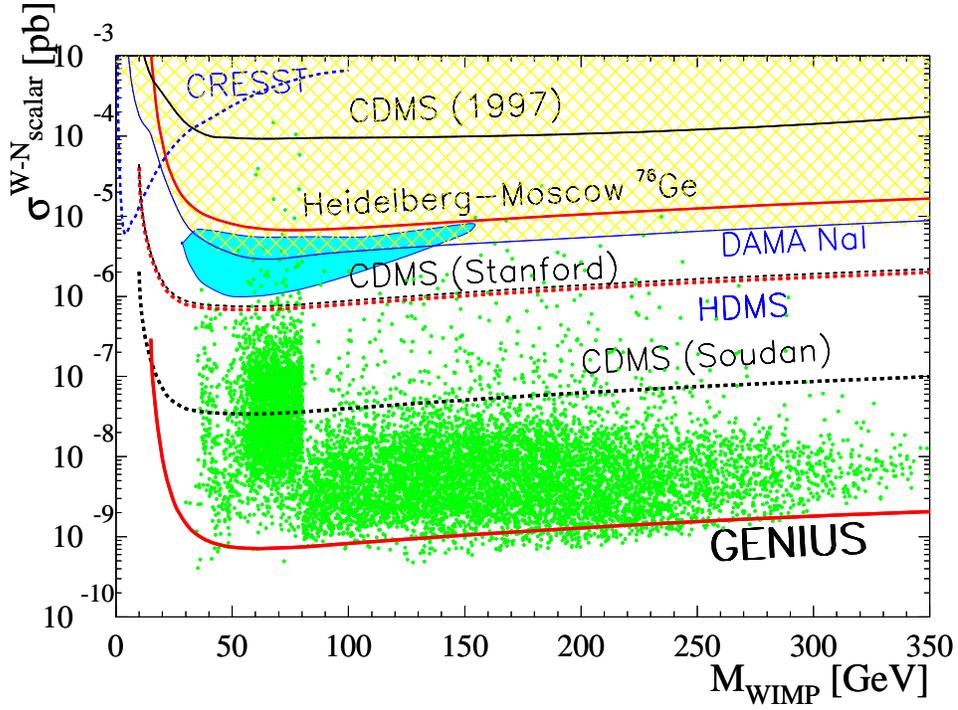}}
\caption{WIMP--nucleon cross section limits as a function of the WIMP
  mass. The hashed region is excluded by the Heidelberg--Moscow
  \cite{prl} and the DAMA experiment \cite{DAMA}, the dashed lines are 
  expectations for recently started or future experiments, like HDMS
  \cite{hdms97}, CDMS \cite{cdms98} and CRESST \cite{cresst96}. The
  filled contour represents  the 2$\sigma$ evidence region of the DAMA
  experiment \cite{damaevid}. The solid
  line is the expectation for the GENIUS project with a background 
  level of 0.01 counts/(keV kg y) and an energy threshold of 11 keV.
The experimental limits are compared to
expectations (scatter plot) for WIMP--neutralinos calculated in the
MSSM framework with non--universal scalar mass unification \cite{bedny1}.}
\label{limits}
\end{figure}

The count rate below 11 keV is dominated by X--rays from 
the decays of $^{68}$Ge, $^{49}$V,     
$^{55}$Fe and $^{65}$Zn (see table \ref{ge_cosmo}).        
Due to their strong contribution, the energy threshold of GENIUS would be
at 11 keV, which is still acceptable (as can be seen from figure \ref{limits}).

\begin{figure}[h]
\epsfxsize=12cm
\centerline{\epsffile{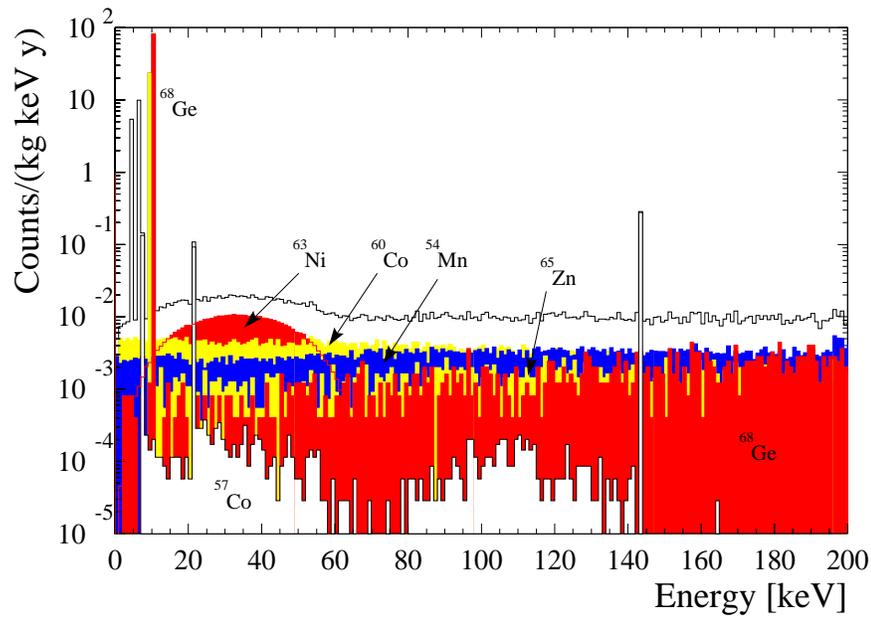}}
\caption{Background originating from cosmic activation of the Ge
  crystals at sea level with 10 days exposure and 3 years deactivation.} 
\label{cosmo}
\end{figure}

Between 11 keV and 70 keV the contribution from $^{63}$Ni dominates
due to the low Q--value (66.95 keV) of the $\beta^-$--decay. 
Figure \ref{cosmo} shows the sum and the single contributions from the
different isotopes.
$^{68}$Ge plays a special role. Since it can not be
extracted by zone melting like all other, non--germanium isotopes,
the starting activity would be in equilibrium with the production
rate. With a half--life of 288 d it would by far dominate the other background
components. A solution could be to process the germanium ore directly
in the underground or to use high purity germanium which
has already been stored for several years in an underground laboratory.    

The sum of all contributions from the cosmogenic activation of the Ge
crystals is 1.9$\times$10$^{-2}$ counts/(kg y keV) between 11 -- 100
keV, for an activation time of 10 days at the Earths surface and a
deactivation time of three years.  
Since this will be the dominant background component in the
low energy region, special attention to short crystal exposure times at
sea level is essential.    
The best solution would be to produce the
detectors underground and to apply strong shielding during the transportation.    

\begin{figure}[h!]
\epsfxsize12cm
\centerline{\epsffile{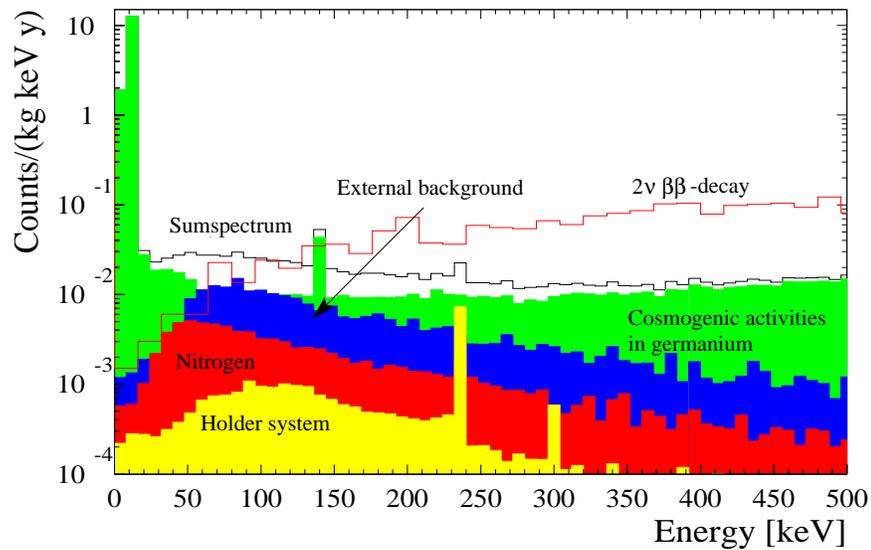}}
\caption{Simulated spectra of the dominant background sources for
a nitrogen tank of 12 m diameter.
Shown are the contributions from the tank walls, the detector holder
system, from neutron capture in the nitrogen,
from natural
radioactivity and from the $^{222}$Rn contamination of the nitrogen.
The solid line represents the sum spectrum of all the simulated
components (note the different channel binning compared to figure \ref{cosmo}).}
\label{specall}
\end{figure}

The two--neutrino accompanied double beta decay of $^{76}$Ge is not
negligible in spite of the low abundance (7.8\%) of this isotope in
natural germanium. The contribution to the background after three
years of measurement is shown in figure \ref{specall}. Due to the
already high statistics reached in the Heidelberg--Moscow experiment
\cite{hdmo}, the half life and spectral form of the decay are well
known and a subtraction of this part raises no difficulties. 
The statistical error of the subtraction is not shown in figure
\ref{limits}.

The cosmogenic activation of the Ge crystals in the Gran Sasso laboratory
is negligible in comparison to the assumed activation scenario at sea level. 
 
\subsubsection{Cosmic activation of the nitrogen at sea level}

An estimation of the production rates of long--lived
isotopes in the nitrogen at sea level reveals the importance
of $^7$Be, $^{10}$Be, $^{14}$C and $^3$H.
The neutron flux at sea level is 8.2$\times$10$^{-3}$cm$^{-2}$s$^{-1}$ 
for neutron energies between 80 MeV and 300 MeV \cite{allkofer}.
Since we didn't find measurements of the cross sections of neutron
induced spallation reactions in nitrogen, 
we assumed that at high neutron energies
(10$^2$--10$^4$ MeV) the cross sections are similar to the proton induced ones.
For the reaction $^{14}$N(n,t$\alpha$n)$^7$Be the cross section was
measured to be (9.0$\pm$2.1) mb at E$_p$ = 450 MeV, (9.3$\pm$2.1) mb at
E$_p$ = 3000 MeV by \cite{reyss} and (13.3$\pm$1.3) mb at E$_p$ = 1600 MeV
by \cite{michel}.
For the reaction $^{14}$N(n,$\alpha$p)$^{10}$Be
the measured cross sections are (1.5$\pm$0.4) mb at E$_p$ = 450 MeV,  
(2.6$\pm$0.6) mb at E$_p$ = 3000 MeV \cite{reyss} and (1.75$\pm$0.11)
mb at E$_p$ = 1600 MeV \cite{michel}.
 
Taking 10 mb for the $^7$Be channel we obtain a production 
rate of 3.3$\times$10$^9$d$^{-1}$ in the whole tank. 
This corresponds with a realistic 10 days sea level exposure after
production by fractional distillation to 4$\times$10$^8$ decays per day.
The simulation of this activity yields a count rate of about
10 events/(kg y keV) in the energy region between 0 -- 100 keV.
This is three orders of magnitude higher than the required level.
However, it can be expected, that a large fraction of $^7$Be is removed
from the liquid nitrogen at the cleaning process for Rn and that in
addition by underground storage (T$_{1/2}$=53.3 d) the contribution of 
$^7$Be is reduced to less than 10$^{-2}$ events/(kg y keV).

For $^{10}$Be, with $\sigma$= 2 mb, the production rate is
6.6$\times$10$^{8}$d$^{-1}$, which is negligible due to the long
half life of T$_{1/2}$=1.6$\times$10$^6$ y.

Tritium in nitrogen can be produced in the following reactions: 
$^{14}$N(n,t)$^{12}$C, $^{14}$N(n,t2$\alpha$)$^{4}$He,
$^{14}$N(n,t$\alpha$n)$^{7}$Be and $^{14}$N(n,tn)$^{11}$C.
The cross section for the production by $^{14}$N(n,t)$^{12}$C was
measured to be 40 mb \cite{Kincaid}. For a rough estimation, we
assumed the same cross section for the other reaction as for the
production of $^{7}$Be to be 10 mb. The total production rate of tritium
corresponds to 2.3$\times$10$^{10}$d$^{-1}$. With T$_{1/2}$=12.33 y, the
activity after 10 days exposure at sea level would be 
3.5$\times$10$^{7}$d$^{-1}$. We simulated 10$^{10}$ decays randomly
distributed in the nitrogen tank. No events were
detected mainly due to the absorption in the dead layer of the p--type 
Ge detectors.

The muon flux at sea level is 1.6$\times$10$^7$m$^{-2}$d$^{-1}$.
The only long--lived isotopes which are produced by inelastic muon
scattering are $^{13}$N and $^{10}$C, with a production rate of about  
3.7$\times$10$^7$ atoms/days (taking $\sigma$ = 500 $\mu$b for both reactions). 
However, $^{13}$N and $^{10}$C are of no relevance due to the
short half lifes of 9.96 m and 19.3 s, respectively. 

The isotopes produced through negative muon capture with half lifes $>$
1s are  $^{14}$C, $^{10}$Be, $^{11}$C and $^{10}$C (see also table
\ref{myonspall}). Again, the number of $^{11}$C and $^{10}$C  
isotopes are soon reduced to a negligible level due to their short
half lifes. The production rate for the whole tank for $^{14}$C is
8$\times$10$^{6}$d$^{-1}$ and 4$\times$10$^{5}$d$^{-1}$ for
$^{10}$Be, which have to be added to the production rates by neutron
capture or spallation reactions.

For the production of $^{14}$C due to the $^{14}$N(n,p)$^{14}$C capture reaction,
three neutron sources at sea level are relevant.
The flux of secondary cosmic ray neutrons with energies between a few
keV and 20 MeV is 
about 2$\times$10$^{-2}$cm$^{-2}$s$^{-1}$\cite{allkofer}. These neutrons penetrate
the wall of the transportation tank and are captured in the liquid 
nitrogen. For a tank surface of 678 m$^2$, about
1.3$\times$10$^5$ s$^{-1}$ neutrons are expected.
The second component are neutrons produced in fast neutron 
spallation reactions in the liquid nitrogen. The production rate of these
neutrons is 2$\times$10$^{4}$ s$^{-1}$ in the nitrogen tank
\cite{lal}. The third component are neutrons produced in muon
reactions, which correspond to 0.85$\times$10$^{3}$ s$^{-1}$ \cite{lal}.
Thus the total flux at sea level is about 1.5$\times$10$^{4}$ s$^{-1}$.
Assuming that every neutron is captured in the nitrogen, yielding
a $^{14}$C nucleus, the production rate of $^{14}$C is about 
1.3$\times$10$^{10}$d$^{-1}$. For a production and transportation 
time of ten days, the simulation yields
less than 10$^{-4}$ counts/(kg keV y) in the relevant energy
region. Through the purification of the nitrogen this
contribution will be further reduced.

\subsection{Sum spectrum from simulations}

In table \ref{backlist} the components discussed so far are listed
and summed up. 
Not included in the table are the contributions from the intrinsic
impurities in the Ge crystals and from the $^{7}$Be activation of the
liquid nitrogen during its transportation at sea level. 
For the Ge--crystals we have only very conservative upper values for
their true contamination, which is expected to be much lower (see
Section 3.2.1).
Regarding the $^{7}$Be contamination of liquid nitrogen, we are
confident that the cleaning efficiency is high enough in order to
reduce this contribution to a negligible level. 
Assuming a background as stated above, we will achieve
a mean count rate in the interesting
region of about 3.1$\times$10$^{-2}$ events/(kg y keV). This
means a further reduction of background in comparison to our 
best measurement (about 20 counts/(kg y keV) below 100 keV \cite{prl}) 
by about 3 orders of magnitude.

In figure \ref{specall} the spectra of individual contributions 
and the summed up total background spectrum are shown. 
As mentioned before, the low energy spectrum is dominated by events
originating from the cosmogenic activation of the Ge crystals at the
Earths surface. Production of the detectors underground would
significantly reduce this contribution.  

\section{Conclusion and Outlook}

After a first study of the dominant background sources in the GENIUS
experiment \cite{hellmig97,hellklap}, we performed a more detailed
investigation of the background in the energy region relevant for the 
direct detection of WIMPs.  At the same time  a technical study with
three naked Ge crystals in a common crystal holder immersed in liquid 
nitrogen was carried out.

We simulated an exact tank design with five natural germanium
detectors supported by a holder system of high molecular polyethylene.
The measured photon flux in the Gran Sasso laboratory with
photon energies between 0 and 3 MeV was simulated for all the
dominant radioisotopes. The influence of neutrons from
the natural radioactivity of the rocks was estimated under the
conservative assumption that every neutron reaching the liquid shielding
is captured leading to excited $^{14}$C$^*$ and $^{15}$N$^*$ nuclei.
We simulated the radon contamination of the nitrogen and
studied the activation of the shielding by muon induced
nuclear disintegration and by secondary neutrons from muon
interactions. At the same time we calculated the activation
of the Ge crystals and of the liquid nitrogen during the period of
their exposure at sea level. 

The obtained count rates and activation levels confirm
the possibility to achieve a background count rate
of 10$^{-2}$ counts/kg keV y in the energy region below 100 keV.

For this, we need only fairly standard 
arrangements for the setup, like  
 a boron implanted polyethylene foam isolation or a neutron absorption 
film, an anticoincidence shield for muons (scintillators on top of the
setup) and a nitrogen-cleaning device (eventually with nitrogen recycling).

Besides that, we need more sensitive measurements of the contamination
level of materials to hold the Ge-crystals and the Rn contamination of liquid nitrogen.
Both measurements are underway.
Furthermore, surface contamination of the crystals and the supporting
structure needs high attention.

Reaching the background level aimed at, the GENIUS project could bring
a large decisive
progress in the field of direct dark matter search. 
It could probe 
a major part of the SUSY--WIMP parameter space interesting for the detection of
neutralinos, thus possibly deciding whether or not neutralinos are the major
component of the dark matter in our Galaxy. 

\section*{Acknowledgments}
L.B. is supported by the Graduiertenkolleg of the University of
Heidelberg. She would like to thank Y. Declais for helpful discussions.

\newpage

\begin{table}[htb]
\begin{center}
\setcounter{mpfootnote}{0}
\vskip0.3cm
\centering
\begin{tabular}{lcc}
\hline
Isotope & Energy [keV]&Flux [m$^{-2}$d$^{-1}$]\\
\hline
$^{40}$K   & 1460    & 3.8$\times10^{7}$\\
\hline
$^{214}$Pb & 295.2   & 0.8$\times10^{7}$\\
$^{214}$Pb & 352     & 1.8$\times10^{7}$\\
$^{214}$Bi & 609.3   & 2.9$\times10^{7}$\\
$^{214}$Bi & 1120.3  & 1.4$\times10^{7}$\\
$^{214}$Bi & 1764.5  & 1.7$\times10^{7}$\\
\hline
$^{208}$Tl & 2614.5  & 1.35$\times10^{7}$\\
\hline
\end{tabular}
\caption{Simulated components of the gamma ray flux from natural
  radioactivity in the Gran Sasso Laboratory (from \cite{arpesella92}).}
\label{photonGS}
\end{center}
\end{table}

\begin{table}[htb]
\vskip0.3cm
\centering
\begin{tabular}{lcc}
\hline
Reaction & T$_{1/2}$ of the product& Decay energy\\
\hline
$^{14}$N(n,p)$^{14}$C & T$_{1/2}$=5.7$\times$10$^3$y& E$_{{\beta}^-}$=0.16 MeV\\
$^{14}$N(n,$\gamma$)$^{15}$N & stable& \\
$^{14}$N(n,2n)$^{13}$N & T$_{1/2}$= 9.96 m& E$_{{\beta}^+}$=1.2 MeV\\
$^{14}$N(n,$\alpha$)$^{11}$B  & stable&\\
$^{14}$N(n,t)$^{12}$C & stable&\\
$^{14}$N(n,2$\alpha$)$^{7}$Li &stable&\\
\hline
\end{tabular}
\caption{Neutron interactions in the liquid nitrogen for neutron
  energies $<$ 20 MeV.}
\label{neutrons}
\end{table}

\begin{table}[htb]
\vskip0.3cm
\centering
\begin{tabular}{lccc}
\hline
Reaction & T$_{1/2}$ & Decay energy & Rate [y$^{-1}$]\\
\hline
$^{14}$N($\mu$,$\nu_{\mu}$)$^{14}$C & T$_{1/2}$=5.7$\times$10$^4$y& E$_{{\beta}^-}$=0.16 MeV&584\\
$^{14}$N($\mu$,$\nu_{\mu}\alpha$)$^{10}$Be & T$_{1/2}$=1.6$\times$10$^{10}$y& E$_{{\beta}^-}$=0.6 MeV&29\\
$^{14}$N($\mu$,$\nu_{\mu}$p)$^{13}$B & T$_{1/2}$=17.33ms& E$_{{\beta}^-}$=13.4 MeV&116\\
$^{14}$N($\mu$,$\nu_{\mu}$n)$^{13}$C &  stable&&3798\\
$^{14}$N($\mu$,$\nu_{\mu}\alpha$n)$^{9}$Be & stable&&17\\
$^{14}$N($\mu$,$\nu_{\mu}\alpha$p)$^{9}$Li & T$_{1/2}$=178ms& E$_{{\beta}^-}$=13.6 MeV&0.6\\
$^{14}$N($\mu$,$\nu_{\mu}$2n)$^{12}$C & stable&&1168\\
$^{14}$N($\mu$,$\nu_{\mu}$3n)$^{11}$C & T$_{1/2}$=20.38m&
E$_{{\beta}^-}$=13.4 MeV, E$_{\gamma}$=4.44 MeV&292\\
$^{14}$N($\mu$,$\nu_{\mu}$4n)$^{10}$C & T$_{1/2}$=19.3s& E$_{{\beta}^+}$=1.9 MeV&117\\
\hline
\end{tabular}
\caption{Spallation reactions from muon capture.}
\label{myonspall}
\end{table}

\begin{table}[htb]
\begin{center}
\setcounter{mpfootnote}{0}
\vskip0.3cm
\centering
\begin{tabular}{lcc}
\hline
Muon flux & 124 h$^{-1}$&\\
\hline
Stopped muons & 86 d$^{-1}$&\\
\hline
Decaying muons& $\mu ^+$ & $\mu ^-$\\
& 50  d$^{-1}$& 20 d$^{-1}$\\
\hline
Captured muons &$\mu ^+$ & $\mu ^-$\\
& 0 & 16 d$^{-1}$\\
\hline
\end{tabular}
\caption{Muon flux, stopped, captured and decaying muons in the
  nitrogen shielding of the Genius detector.}
\label{mcapture}
\end{center}
\end{table}

\begin{table}[htb]
\begin{center}
\begin{tabular}{lcc}
\hline
Source & Radionuclide & Purity \\
\hline
Nitrogen & $^{238}$U  & 1$\times$10$^{-15}$g/g \\
           & $^{232}$Th & 5$\times$10$^{-15}$g/g \\
           & $^{40}$K   & 1$\times$10$^{-15}$g/g \\
\hline
Steel vessel  & U/Th       & 1$\times$10$^{-8}$g/g \\
\hline
\end{tabular}
\caption{Restrictions in radioactive purity of single components for
the liquid nitrogen.  }
\label{forderung}
\end{center}
\end{table}

\begin{table}[htb]
\begin{center}
\setcounter{mpfootnote}{0}
\vskip0.3cm
\centering
\begin{tabular}{lccc}
\hline
Isotope & Decay mode, T$_{1/2}$ & Energy [keV]& Activity [$\mu$Bq kg$^{-1}$]\\
\hline
$^{49}$V   &  EC, 330 d    & no $\gamma$, E (K$_{\alpha}$ $^{49}$Ti)=4.5 & 0.17\\
$^{54}$Mn  &  EC, 312.2 d    &E$_{\gamma}$=1377.1 E (K$_{\alpha}$ $^{54}$Cr)=5.99   & 0.20\\
$^{55}$Fe  &  EC, 2.73 a    & no $\gamma$, E (K$_{\alpha}$ $^{55}$Mn)=5.9   & 0.31\\
$^{57}$Co  &  EC, 271.3 d  & 136.5 (99.82\%)E (K$_{\alpha}$ $^{57}$Fe)=7.1  & 0.18\\
$^{60}$Co  &  $\beta ^-$, 5.27 a & 318 (99.88\%), E$_{\gamma 1,2}$=1173.24, 1332.5&0.18\\
$^{63}$Ni & $\beta^-$, 100.1 a & E$_{\beta^-}$= 66.95 no $_{\gamma}$ & 1.2$\times$10$^{-2}$\\
$^{65}$Zn & EC, 244.3 d & E$_{\gamma}$=1115.55 (50.6\%),E (K$_{\alpha}$ $^{65}$Cu)=8.9 & 1.14\\
$^{68}$Ge & EC, 288 d &no $\gamma$, E (K$_{\alpha}$ $^{68}$Ga)=10.37,
Q$_{EC}$($^{68}$Ga)= 2921 &2.56\\
\hline
\end{tabular}
\caption{Cosmogenic produced isotopes in the Ge crystals for an
  exposure time at sea level of 10 days and for 3 years deactivation time.}
\label{ge_cosmo}
\end{center}
\end{table}

\renewcommand{\baselinestretch}{1}{

\begin{table}[htb]
\begin{center}
\begin{tabular}{lcc}
\hline
Source & Component & Count rate (11-100 keV) [counts/(kg y keV)] \\
\hline
Nitrogen   & $^{238}$U   &  2$\times$10$^{-3}$ \\
intrinsic  & $^{232}$Th  &  1$\times$10$^{-3}$ \\
           & $^{40}$K    &  1$\times$10$^{-4}$ \\
           & $^{222}$Rn  &  1$\times$10$^{-3}$ \\
N activation & $^{14}$C    &  1$\times$10$^{-4}$ \\
\hline
Steel vessel   & U/Th        &  1.5$\times$10$^{-5}$ \\
\hline
Holder system  & U/Th        &  8$\times$10$^{-4}$ \\
\hline
Surrounding    & Gammas      &  4$\times$10$^{-3}$\\
               & Neutrons    &  4$\times$10$^{-4}$\\
               & Muon shower &  2$\times$10$^{-3}$ \\
               & $\mu$ $\rightarrow$ n       &  1$\times$10$^{-3}$ \\
               & $\mu$ $\rightarrow$ capture & $<<$1$\times$10$^{-4}$ \\ 
\hline
Cosmogenic     & $^{54}$Mn   & 3$\times10^{-3}$\\
activities     & $^{57}$Co   & 1$\times10^{-3}$\\
in the crystals    & $^{60}$Co   & 4$\times10^{-3}$\\
               & $^{63}$Ni   & 8$\times10^{-3}$\\
               & $^{65}$Zn   & 2$\times10^{-3}$\\
               & $^{68}$Ge   & 1$\times10^{-3}$\\
\hline

Total          & &  3.1$\times$10$^{-2}$ counts/(kg y keV) \\
\hline 
\end{tabular}
\end{center}
\caption{Summation of background components in the region 0 keV--100 keV.} 
\label{backlist}
\end{table}
}

\end{document}